\documentclass[conference]{IEEEtran}
\usepackage{blindtext, graphicx}
\usepackage[sorting=none]{biblatex}

\begin{document}

\title{A Comparative Survey of LPWA Networking}

\author{\IEEEauthorblockN{Joseph Finnegan}
\IEEEauthorblockA{Department of Computer Science\\
Maynooth University\\
Maynooth, Ireland\\
Email: joseph.finnegan@cs.nuim.ie}
\and
\IEEEauthorblockN{Stephen Brown}
\IEEEauthorblockA{Department of Computer Science\\
Maynooth University\\
Maynooth, Ireland\\
Email: stephen.brown@nuim.ie}}

\maketitle

\begin{abstract}

Motivated by the increasing variance of suggested Internet of Things (IoT) applications and the lack of suitability of current wireless technologies in scalable, long range deployments, a number of diverging Low Power Wide Area (LPWA) technologies have been developed. These technologies promise to enable a scalable high range network on cheap low power devices, facilitating the development of a ubiquitous IoT. This paper provides a definition of this new LPWA paradigm, presents a systematic approach to defined suitable use cases, and undertakes a detailed comparison of current LPWA standards, including the primary technologies, upcoming cellular options, and remaining proprietary solutions.

\end{abstract}


\IEEEpeerreviewmaketitle

\section{Introduction}

The use of wireless communications has become ubiquitous in everyday life. The ongoing development of the IoT will only expand this. IoT devices are characterised by the shift in the role of human interaction in the device's run cycle; IoT devices will be autonomous, embedded in the world around us, collecting data and providing services. The shift will be away from human-generated data and human-requested services, and towards machine-generated data and notification-based services. Information will be sensed and data generated without human interaction, enabling the automation of previously monotonous tasks by independent networks of devices \cite{Gubbi}. Cisco predicts there will be 12.2 billion connected devices by 2020 \cite{Cisco}, and the EU predicts 6 billion IoT connections within the EU by 2020 \cite{EU}. The nature of the communication of these devices will differ from current human-controlled devices (e.g. mobile phones and laptops); the most notable difference will be that uplink communication will take up a higher percentage of traffic - they will be data \textit{producing} rather than consuming \cite{Shafiq}. 




In terms of communication requirements, until recently the majority of devices have been members of a small set of types of devices: stationary, mains-powered devices such as personal computers and printers that can use Ethernet, mobile devices with rechargeable energy supplies such as mobile phones and laptops which can use Wi-Fi and/or cellular, and some low power wireless devices that perform set single tasks. In the near future, a much wider variance in Internet of Things devices is expected \cite{Margelis}. IoT devices will be used in the smart home, smart city, in industrial applications, agricultural applications, and monitoring and sensing applications.

Some of the available wireless protocol options for IoT devices are Bluetooth LE, 802.15.4-based Zigbee, 802.11ah-based HaLoW, Wi-Fi, and cellular. However, these options are limited in that they cannot easily provide long range communication in devices that must operate at low power. This has motivated the development of a number of new wireless protocols designed specifically for long distance, low power devices, which have been designated LPWA. Depending on the specific requirements of the application the particular Quality of Service (QoS) requirements will change, and so the particular wireless technology which is most suited to the scenario will depend on the specific details of the application. In planning, a trade-off must be made by the designer in choosing a technology that fits all of the requirements in terms of the key metrics for IoT devices; that is, range, energy, throughput, and cost \cite{Samie} - there is no one size fits all solution. 




The goal of this paper is to provide a clear definition of LPWA and to differentiate LPWA from other technologies in terms of these key metrics. Use cases are then discussed motivated by these metrics; each wireless technology has different advantages and disadvantages, and if LPWA technologies are to gain traction it is important that they are to be applied in the correct application areas. Finally, existing LPWA technologies are discussed and compared. This research differs from other LPWA-focused surveys \cite{Samie, Andreev, Elkhodr, Raza, Sanchez-Iborra, Ali, Moyer, Song, Guibene} in that the scope of LPWA  has been broadened to include incoming cellular-based technologies, namely NB-IoT and EC-GSM-IoT, as well as proprietary solutions, and in that a more detailed direct comparison of the available technologies has been performed.  


\section{Definition of LPWA}

LPWA technologies are characterised by their focus on energy efficiency, scalability, and coverage \cite{Raza}. These technologies typically operate in the unlicensed sub 1GHz Industrial, Scientific and Medical (ISM) band, which is shared by all ``Short Range'' devices (Short Range devices being defined by ETSI \cite{ETSI} as devices at low risk of interference; including alarms, identification systems, radio-determination, telecommand, telemetry, RFID, and detection, movement and alert applications). Zwave, 802.11ah, 802.15.4g, and 802.15.4 are examples of protocols that also make use of this band \cite{Andreev}. The majority of LPWA technologies can be separated into either wideband or ultra-narrowband technologies, where wideband techniques utilise a larger bandwidth than what is needed and use controlled frequency diversity to retrieve data, and ultra-narrow band techniques compress data into ultra narrow bands and use high stability RF crystals and digital signal processing techniques to recover the data \cite{Reynders, Goursaud}.  

Because of the popularity of this band, and as there are multiple protocols running in the same band simultaneously, these devices are subject to regulations, which vary by region but typically either necessitate the use of Listen-Before-Talk (LBT) mechanisms or limit the devices from communicating more than a particular percentage of the time during a day, fundamentally limiting the potential throughput of the device \cite{ETSI}. In Europe, the ETSI regulations define subbands for use within the ISM bands. As an example, Table I outlines the spectrum access available in the 868MHz band for subbands that support wideband modulation. To minimise bandwidth spent on the control plane of the network, LPWA technologies usually operate on a star network topology, where nodes communicate directly with a base station. As the base stations themselves are also subject to the same duty cycle regulations, LPWA protocols are focused on applications that do not require large amounts of data being relayed from the gateway out to the nodes (i.e. minimal downlink traffic) \cite{Andreev} and applications that do not report critical information with high reliability requirements (as use of acknowledgement packets is limited). LPWA technologies address the restrictions of cellular, namely high power consumption and high cost of service, and enable long range communication on autonomous low power devices \cite{Mahmoud}. Applications that particularly meet the needs for LPWA are those that require autonomous battery-powered nodes with a long range, a long network lifetime, low throughput, and do not have strict requirements on latency.

\begin{table}[h]
\centering
\caption{ETSI Spectrum Access per subband}
\label{my-label-5}
\begin{tabular}{|l|l|l|}
\hline
Subband         & Spectrum Access    & Edge Frequencies \\ \hline
g               & 1  \% or LBT AFA   & 865-868MHz       \\ \hline
g               & 0.1\% or LBT AFA   & 865-870MHz       \\ \hline
g1              & 1  \% or LBT AFA   & 868-868.6MHz     \\ \hline
g2              & 0.1\% or LBT AFA   & 868.7-869.2MHz   \\ \hline
g3              & 10 \% or LBT AFA   & 869.4-869.65MHz  \\ \hline
g4              & No Requirement     & 869.7-870MHz     \\ \hline
g4              & 1  \% or LBT AFA   & 869.7-870MHz     \\ \hline
\end{tabular}
\end{table}

\begin{figure}[h]
  \centering
  
    \includegraphics[scale=0.6]{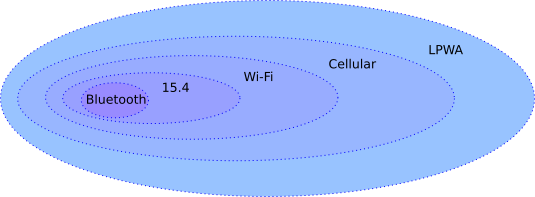}
    \caption{Typical Range of Wireless Technologies}
\end{figure}

Low Power in the context of this subspace of wireless technologies refers to a level of power usage in a typical application would can enable the node to function autonomously for an extended period of time using only a cheap power source, such as a pair of AA batteries or a button cell. 

Wide Area refers to a distance that two nodes can communicate directly that is large enough to enable the coverage of a city or rural area without an unreasonably complex mesh network of nodes. The applications that best fit the use of LPWA technologies either do not or cannot easily replace their power source - if they could then the use of a higher power technology would in most cases be a more suitable solution. 

Currently, there are a number of competing LPWA technologies, all of which differ in specific details but fundamentally focus on the same target market. The majority of this paper will focus on a detailed description and comparison of the more open LPWA standards: LoRa, Sigfox, NB-IoT, and EC-GSM-IoT. Other competing standards will also be discussed, namely Nwave, Telensa, Weightless-P, RPMA (Ingenu), the Dash 7 Alliance Protocol, and NB-Fi (WAVIoT).

\section{Use Cases of LPWA} 


In previous research, LPWA use cases have been described and defined based on potential applications areas, using imprecise terms (e.g. with ``sport'' or ``health'' chosen as use cases). We believe a systematic approach is required. IoT wireless protocol options should be defined in terms of the 3 key metrics (range, throughput, and energy efficiency), and from this potential use cases can be categorised and matched to suitable wireless protocols. In Figure 2 we define 8 categories based on the key metrics, and give examples of protocols that match those subsections.  


\begin{figure}[h]
  \centering  
    \includegraphics[scale=0.6]{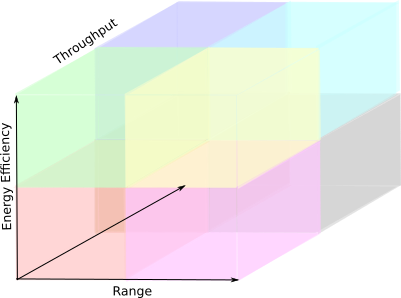}
    \caption{Wireless Technology categories}
\end{figure}

\begin{table}[h]
\centering
\label{my-label-4}
\begin{tabular}{|l|l|l|l|}
\hline
Range    & Throughput     & Energy Efficiency       &    Example Protocol    \\ \hline
Low      & Low            & Low                     &    ---                 \\ \hline
Low      & Low            & High                    &    Bluetooth LE        \\ \hline
Low      & High           & Low                     &    802.11n             \\ \hline
Low      & High           & High                    &                        \\ \hline
High     & Low            & Low                     &                        \\ \hline
High     & Low            & High                    &    LoRaWAN (LPWA)      \\ \hline
High     & High           & Low                     &    LTE                 \\ \hline
High     & High           & High                    &                        \\ \hline
\end{tabular}
\end{table}

\begin{table*}[h]
\centering
\caption{Use Cases \& requirements}
\label{my-label-4344}
\begin{tabular}{|l|l|l|l|l|l|}
\hline
Use Case                    & Range         & Throughput & Energy Efficiency        & LPWA Suitability   & Reference      \\ \hline

Air quality management      & High          & Low        & High                     & X         & \cite{Margelis}  \\ \hline
Water metering              & High          & Low        & High                     & X         & \cite{Margelis} \\ \hline
Power metering              & High          & Low        & High                     & X         & \cite{Margelis} \\ \hline
Smart waste management      & High          & Low        & High                     & X         & \cite{Margelis} \\ \hline
Natural gas usage           & High          & Low        & High                     & X         & \cite{Margelis}  \\ \hline

Building automation         & Medium        & Low        & Medium                   & X         & \cite{Song} \\ \hline
Heating control             & Medium        & Low        & Medium                   & X         & \cite{Samie}\\ \hline
Access control              & Medium        & Low        & Medium                   & X         & \cite{Samie} \\ \hline
Structural health monitoring & High         & Low        & High                     & X         & \cite{Samie} \\ \hline

Soil monitoring             & High          & Low        & High                     & X         & \cite{Adelantado} \\ \hline
Crop growth monitoring      & High          & Low        & High                     & X         & \cite{Adelantado} \\ \hline
Livestock monitoring        & High          & Low        & High                     & X         & \cite{Adelantado} \\ \hline

Fleet management            & High          & Medium     & Low                      &           & \cite{Petajajarvi} \\ \hline
Asset tracking              & High          & Low        & High                     & X         & \cite{Petajajarvi} \\ \hline
Usage-based insurance       & High          & Medium     & Low                      &           & \cite{Petajajarvi} \\ \hline
Smart traffic               & Medium        & Medium     & Low                      &           & \cite{Petajajarvi} \\ \hline
Parking management          & Medium        & Medium     & High                     &           & \cite{Margelis} \\ \hline
On-street lighting control  & High          & Low        & High                     & X         & \cite{Song} \\ \hline
Self service bike rentals   & Medium        & Medium     & Low                      &           & \cite{Margelis}  \\ \hline

Wearable devices            & Low           & High       & Low                      &           & \cite{Petajajarvi-2} \\ \hline            
\end{tabular}
\end{table*}

LPWA technologies are particularly suited for pervasive IoT applications that are needed to generate a predictable, small amount of data over long periods; for applications that require the transfer of little data over large distances i.e. have strong range and energy requirements, but not strong throughput requirements. In terms of smart cities, this covers use cases such \textit{parking management}, \textit{air quality monitoring}, \textit{water metering}, \textit{power metering}, \textit{smart waste management}, \textit{natural gas usage}, \textit{on-street lighting control} and \textit{self service bike rentals} \cite{Margelis, Samie, Filho,  Bor, Adelantado}. Use of LPWA technologies instead of mesh-based 802.15.4 protocols in these cases would require less repeaters, less control plane communication, and simpler routing protocols \cite{Filho}. LoRaWAN has already been deployed as both a wireless platform and backend infrastructure in the city of Antwerp, Belgium \cite{Latre}. However, for more time-critical monitoring and metering applications such as wireless industrial control, LPWA technologies cannot guarantee the availability and latency required for the necessary performance \cite{Raza}. Because of the gateway-agnostic approach taken by most LPWA solutions, LPWA technologies are also suitable for mobile applications such as \textit{fleet management}, \textit{asset tracking}, \textit{usage-based insurance} and \textit{smart traffic} \cite{Petajajarvi}. Several researchers have suggested the potential use of LPWA technologies in mobile applications such as remotely piloted aircraft \cite{Kainrath}, Search and Rescue \cite{Dariz}, and automotive applications \cite{Samie}, but the limited throughput and latency guarantees available mean that the usefulness of LPWA in these areas depends on the exact particulars of the proposed applications. 

In rural areas, any application that requires high coverage and has limited uplink traffic requirements and latency needs would be suitable for LPWA. This includes non-critical infrastructure monitoring and environmental monitoring \cite{Mikhaylov}. Agricultural monitoring is another particularly suitable application of LPWA technologies \cite{Adelantado}, where the distances to be covered negate the advantages of lower range technologies, and cellular technologies come with a higher cost. Potential use cases are static applications such as \textit{soil moisture and quality monitoring} and \textit{crop growth monitoring}, and mobile applications such as \textit{livestock monitoring}. LPWA technologies can enable the deployment of smart sensors throughout the area without the management of a complex mesh network. The authors in \cite{Pham} believe that LPWA has particular suitability in developing areas such as sub-Saharan Africa, as LPWA networks in comparison to cellular are cheaper and more energy efficient, and less reliant on pre-existing infrastructure which may be limited in these areas.

Though the range provided by LPWA technologies is typically much higher than would be required to cover a building, there are still benefits to using LPWA technologies in smart building applications. The frequencies used by LPWA protocols can penetrate buildings better than higher frequency protocols such as Wi-Fi, enabling a much simpler deployment of sensors throughout even large office blocks. \textit{Building automation} applications, \textit{heating control}, and \textit{access control} \cite{Samie, Petajajarvi, Petajajarvi-2} could conceivably be implemented using LPWA technologies.  


\section{Key LPWA technologies}

This section discusses in detail the main LPWA technologies. Choice of key technologies has been defined based on the amount of data publicly available on each technology, as well as the amount of published papers focused on each and the breadth of applications that each technology targets. 

\subsection{LoRa and LoRaWAN}

LoRa is a physical layer technology developed by Semtech which makes use of spread spectrum modulation, which leads to a higher link budget and better resilience against interference \cite{Augustin}. LoRa is effectively a form of Chirp Spread Spectrum (CSS) with integrated Forward Error Correction (FEC) \cite{Bor}. In typical use cases, LoRa communicates over the license-free sub-1GHz ISM bands. In Europe, 433MHz and 868MHz are available, with 868MHz being most commonly used as it is broader and contains subbands with less strict duty cycle requirements. 

LoRa is proprietary and owned by Semtech, but the upper layers of the network stack are open for development \cite{Augustin}. The most well supported upper layer protocol for LoRa is LoRaWAN, which is open and managed by the LoRa Alliance, a non-profit. LoRaWAN functions on an operator-free, subscription fee-free model, simplifying and cheapening the management of infrastructure, and helping support a simple ``out-of-the-box'' deployment \cite{Pham-2}. Deploying a LoRaWAN network requires a NetID issued from the LoRa Alliance, or alternatively paying for use through a network provider with a private LoRaWAN network. Alternative upper layer protocols utilising the LoRa physical layer are at the present time limited to one commercial platform \cite{Symphony-Link} and one mesh-adaption published in academia \cite{Bor}.

LoRaWAN supports secure, mobile, GPS-free bi-directional communication for payloads ranging from 19 to 250 bytes. The LoRaWAN overhead per packet is 12 bytes. LoRa range depends on the link budget, which can be modified through changes in bandwidth, coding scheme, transmission power, carrier frequency, and spread factor \cite{Augustin}. Experiments from the research work carried out in \cite{Petajajarvi} and \cite{Aref} indicate that LoRa can achieve a range of up to 5km in urban environments while still successfully retrieving 85\% of packets, up to 30km range in Line-of-Sight measurements, and a range of up to 8km in rural environments while still successfully retrieving 100\% of packets and utilising one of the higher data rate (and so lower range) configurations.    

The spread factor is the ratio between symbol rate and chip rate. The ratio between symbol and chip rate is \(2^{spread factor}\). Six different spread factors are available (between 7 and 12); increasing the spread factor makes the signal more robust to noise, but decreases the data rate. 

\begin{table}[h]
\centering
\caption{Data rate of each spread factor, ignoring duty cycle limits. Coding rate = 4/5, Payload = 20 bytes}
\label{my-label-4}
\begin{tabular}{|l|l|l|l|}
\hline
Spread Factor    & Bandwidth     & Data Rate       &    Receiver Sensitivity (SX1276)       \\ \hline
7                & 125kHz        & 2.22kbits/s     &    -123dBm                             \\ \hline
8                & 125kHz        & 1.19kbits/s     &    -126dBm                             \\ \hline
9                & 125kHz        & 0.64kbits/s     &    -129dBm                             \\ \hline
10               & 125kHz        & 0.35kbits/s     &    -132dBm                             \\ \hline
11               & 125kHz        & 0.16kbits/s     &    -133dBm                             \\ \hline
12               & 125kHz        & 0.08kbits/s     &    -136dBm                             \\ \hline
\end{tabular}
\end{table}

On a LoRa device the bandwidth can be set from 7.8kHz up to 500kHz, though only 125kHz, 250kHz, and 500kHz are typically used. A higher bandwidth corresponds to a higher data rate, a more interference-resilient link, and a lower sensitivity.

Like every protocol in the ISM bands, LoRaWAN must follow the regulations defined by the region the network is deployed in. In Europe, the LoRaWAN standard specifies three channels which must be implemented in every LoRaWAN network - 868.10-868.225, 868.30-868.425, and 868.50-868.625.  Beyond this, the LoRa network provider may define the channels in which devices can communicate, which can be located in any of the available subbands. Each of the available spread factors are orthogonal (or at least orthogonal enough for LPWA use cases \cite{Reynders-2}), enabling multiple signals to be transmitted on the same channel simultaneously.

In practice, a physical channel is chosen on a pseudo-random basis, based on current duty cycle allowances. The LoRaWAN specification defines an ADR (Adaptive Data Rate) scheme which enables the server to set the spread factor in order to best fit the requirements of each node, maximising the battery life of individual nodes while optimising the overall network capacity \cite{Adelantado}. Indeed, \cite{Bor-2} et. al. find that without the use of an ADR-like scheme to manage interference, LoRa networks are not realistically scalable - using the default LoRaWAN settings statically in their network,  only ~120 nodes could be supported in their simulated test application, in comparison to ~1600 nodes in the same network using an ADR-like scheme. 

LoRaWAN also sets further regulations on top of the previously mentioned duty cycle limits, specifying that a subband cannot be used for the next \textit{TimeOnAir*(1/DutyCycleSubband-1)} seconds after the sending of a message \cite{LoRaWAN}. This is to prevent a node from sending a burst of messages on one particular subband until the duty cycle limit has been reached. This time off subband requirement also has to be followed by the gateway, further motivating the minimisation of downlink traffic. LoRaWAN defines 3 types of devices: Class A, which supports basic device-initiated communication, Class B, which extends Class A to add the ability for the network to ping devices (the device is given receive windows at scheduled times), and Class C, which is similar to Class A but in continuous receive mode when not transmitting.

Once the parameters and channel are chosen, the device can then communicate with the gateway using a simple ALOHA-based protocol. As a higher spread factor corresponds to an increase in number of chips used per symbol, the use of higher spread factors leads to a higher energy usage per packet, meaning that the device lifetime and daily throughput are directly dependent on the distance from the device to the nearest gateway. Table IV displays the time on air of identical packets, using different spread factors.    

\begin{table}[h]
\centering
\caption{LoRa Spread factor vs Packets per day. Coding rate = 4/5, Payload = 20 bytes, Bandwidth = 125kHz}
\label{my-label-3}
\begin{tabular}{|l|l|l|l|}
\hline
SF            & Chips per Symbol & Time transmitting (ms)                              & Packets               \\ \hline
7             & 128              & 71.936                                              & 417                   \\ \hline
8             & 256              & 133.632                                             & 224                   \\ \hline
9             & 512              & 246.784                                             & 121                   \\ \hline
10            & 1024             & 452.608                                             & 66                    \\ \hline
11            & 2048             & 987.136                                             & 30                    \\ \hline
12            & 4096             & 1810.432                                            & 16                    \\ \hline
\end{tabular}
\end{table}

The performance of a LoRaWAN network is limited by the strict access limitations imposed by the regional regulations, the further LoRaWAN regulations, and the limitations of the simple ALOHA-based medium access control, which is not suited for dense and busy networks \cite{Georgiou}. To try alleviate these, the authors in \cite{Pham-2} propose the central management of network access from the gateway, enabling a device to go beyond its designated duty cycle if given permission from the gateway, which ``borrows'' time from other devices in the network. This enables a device to always be able to communicate in high-priority scenarios. However, the legality of this in larger scale LoRa networks is unclear. For now, in not overly dense networks LoRa is limited by the regulations, not by the technical limitations of the physical layer technology \cite{Adelantado, Bor}. 
 
\subsection{Sigfox}

Sigfox's technology \cite{Sigfox} is a proprietary, ultra-narrowband approach, operating on the unlicenced sub 1GHz ISM bands. Sigfox functions on an operator model - users purchase end devices, and instead of deploying gateways, purchase subscriptions for each device to regional Sigfox-supported networks operated by network providers, who manage the network of gateways. The company already claim full coverage across most of western Europe. The system is a cloud-based approach, where all data received by the gateways is sent to a backend server, where it can be accessed by the customer through a web portal. Customers can then implement callbacks to have their data relayed on to their own system \cite{Nolan}. 

Sigfox splits the subband it utilises (868.180MHz to 868.220MHz) into 400 separate 100Hz subbands, forty of which are reserved \cite{Libelium}. The noise level in each of these narrow bands (hence ``ultra-narrowband'') is very low, enabling the simple decoding of signals at the receiver. The receiver is capable of demodulating a very low received power signal (-142dBm). A Sigfox base station can cover a range of 20-50km in rural areas, and 3-10km in urban areas. The sharing of the frequency space in this manner also increases the number of devices that can be supported. However, it also decreases the data rate \cite{Raza}.

The channel access method of Sigfox is effectively R-FDMA (Randomised-FDMA) with no channel pre-transmission sensing (e.g. LBT). When sending an uplink packet, the end device randomly chooses three of the unreserved 360 channels and sends the packet to the base station. The base station scans the spectrum listening at every channel and uses signal processing algorithms to retrieve the message \cite{Margelis}. This redundancy helps to ensure delivery, as the very limited downlink traffic prevents the regular use of acknowledgement messages. 

As Sigfox uses the same frequency subband as LoRa, it also must adhere to the same duty cycle regulations. Sigfox provide different engagement models (number of 12 byte messages allowed per day) for different prices - all of which hold the end devices to the regulations. The platinum level is the most amount of messages that can be sent to still hold to the regulations for the 868 band.

\begin{table}[h]
\centering
\caption{Sigfox throughput}
\label{my-label-6}
\begin{tabular}{|l|l|l|}
\hline
Scheme        & Number of packets       & Max. bytes per day (uplink)    \\ \hline
Platinum      & 101-140 + 4 downlink    & 1680                           \\ \hline
Gold          & 51-100 + 2 downlink     & 1200                           \\ \hline
Silver        & 3-50 + 1 downlink       & 600                            \\ \hline
One           & 1-2 + no downlink       & 24                             \\ \hline
\end{tabular}
\end{table}

As Table V indicates, Sigfox is not a fully bidirectional technology - earlier versions of the technology were uplink only, and now though some minimal amount of downlink traffic is supported per subscription level, downlink messages can only precede uplink messages, after the sending of which the device should wait for a response. Uplink and downlink communication also utilise different modulation schemes: for uplink, a BPSK scheme operating at a fixed 100bps is used because of its spectral efficiency \cite{Reynders}. For downlink, a GFSK scheme operating at 500bps on a 600Hz spectrum segment is used. As mentioned above, 12 byte payloads are supported for uplink. 8 bytes are supported for downlink. The protocol overhead is 26 bytes.

Like LoRaWAN, the Sigfox MAC layer is based on the simple unslotted ALOHA MAC protocol, the only difference being that Sigfox limits the number of messages an end device can send \cite{Reynders}. This is an energy-efficient approach, as no energy is used for medium sensing and, since time synchronisation is not required, control plane packets are unnecessary \cite{Goursaud}. The user must purchase a subscription for each end device, and the particular level of subscription defines the maximum number of packets that that device can send. 

\subsection{NB-IoT}

Narrowband Internet of Things (NB-IoT) is one of three solutions, along with EC-GSM-IoT and eMTC, forming 3GPP's Cellular-IoT (C-IoT), in anticipation of the development of the Internet of Things \cite{3GPP-tech}. Each of the three were introduced and defined in 3GPP Release 13, and are expected to be further defined and upgraded in 3GPP Release 14. Whereas the other newly defined cellular technologies can be considered advances on previous work (EC-GSM-IoT is designed to enhance GSM, and LTE-MTC will enhance LTE), NB-IoT can be considered a new track, with good co-existance performance but not fully backward compatibility with existing 3GPP technologies \cite{Wang}. eMTC performs at too high of a data rate to be comparable to LPWA technologies; EC-GSM-IoT will be discussed in a later section. A clear distinction between cellular approaches and other LPWA technologies is that these cellular approaches operate on licenced bands and so do not have to deal with the same duty cycle regulations as other technologies.

Essentially, NB-IoT is built from LTE with added simplifications e.g. a modified acquisition process (different cell search process to LTE), reduced bandwidth requirements (using 180kHz of bandwidth, in comparison to 1.4-20MHz used by LTE), a modified random access scheme - resulting in a fast development time. Enhanced coverage and reduced power consumption is achieved in exchange for relaxed latency, a lower data rate, and lower spectrum efficiency. The price of the chip is also reduced, through the use of a narrower band \cite{Nakamura}. Deployment of NB-IoT can be provided through a software update.

NB-IoT supports 3 different deployment scenarios:

\begin{itemize}

\item \textit{In-band operation}: deployed within an LTE wideband system - comprising 1 or more of the LTE Physical Resource Blocks (180kHz). The transmit power at the base station is shared between wideband LTE and NB-IoT, and both technologies can be supported using the same base station hardware, without compromising the performance of either \cite{Wang, Ratasuk-2}.

\item \textit{Standalone}: deployed in a standalone 200kHz of spectrum. All transmission power at the base station is used for NB-IoT, increasing coverage. Typical usage of this mode would be as replacement of GSM carriers.

\item \textit{Guard-band operation}: co-located with an LTE cell, placed in the guard band of an LTE carrier. This shares the same power amplifier as LTE channel, and so shares transmission power \cite{Yu}.

\end{itemize}

The downlink of NB-IoT is based on OFDMA, with 15kHz subcarrier spacing, and reuses the same OFDM numerology as LTE \cite{Wang}. Both single-tone and multi-tone are supported in the uplink. Multi-tone is based on SC-FDMA with 15kHz subcarrier spacing. With single-tone, sub-carrier spacing can be 15kHz or 3.75kHz \cite{Lin}. NB-IoT achieves a 20 dB improvement over GPRS, giving a maximum coupling loss of 164dB \cite{Rico-Alvarino}. 

NB-IoT targets covering 52k devices per channel per cell. This is based on an estimation of 40 devices per household, in an area with the density of London \cite{Rico-Alvarino, Ratasuk}. NB-IoT aims to enable a typical device lifetime of over ten years, on a battery capacity of 5Wh. NB-IoT, like LTE, uses discontinuous reception (DRX), which avoids monitoring the control channel continuously in order to conserve energy. LTE has DRX cycles up to 2.56s. Release 13 introduced extended DRX (eDRX) cycles for both idle and connected modes, which extend the cycles to 43.69 minutes and 10.24 seconds respectively \cite{Rico-Alvarino}, further increasing energy conserved.

LTE-based IoT solutions (including NB-IoT) will have a SIM-like approach - adding an extra cost in the form of a subscription charge. Expected subscription charge information is not yet available.  Deployments of NB-IoT will begin properly in 2017, though preliminary deployments have started at the end of 2016. 

\subsection{EC-GSM-IoT}

EC-GSM-IoT, also known as EC-GSM, is another LPWA technology in development by 3GPP \cite{3GPP-tech}. It is designed as an enhancement to GSM, and re-uses the current GSM design whenever possible, only making changes that are necessary in order to enhance LPWA-related requirements, that is, high capacity, long range, and low energy. The re-use of GSM design means that upgrades to GSM networks can be provided with a software upgrade, and support for new devices can be achieved in existing GSM deployments. In addition, already deployed GSM units will not be adversely effected with the deployment of EC-GSM devices, as traffic from legacy GSM devices and EC-GSM-IoT devices can be multiplexed on the same physical channels - the multiplexing principles from GSM are carried over to EC-GSM-IoT \cite{Nokia}. EC-GSM-IoT uses 200kHz of bandwidth per channel, for a total system bandwidth of 2.4MHz. The first commercial launches are planned for 2017.

On the downlink physical layer, the design is for the most part the same as current GSM. The primary difference is that a new packet control channel format has been designed to limit the amount of control signalling required. On the uplink physical layer, this new control channel format is also used, along with an overlaid CDMA technique (on EC-PDTCH/U, EC-PACCH/U, and on the EC-RACH) to increase capacity, enabling multiple devices to transmit on the same physical channel simultaneously \cite{3GPP-tech}. Beyond this, the design follows GSM principles. There are two random access channels in EC-GSM - if the device needs to provide normal GSM coverage, legacy RACH is used. If the device needs to provide extended coverage, EC-RACH is used. 

Extending the coverage of GSM is achieved through the use of blind repetitions. Different coverage classes are defined, with different numbers of total blind transmissions for different logical channels. The coverage aimed for in EC-GSM is 164 dB MCL \cite{Nokia} for the 33 dBm power class and 154 dB MCL for the 23 dBm power class. 50,000 devices can be supported per cell. The throughput rate of the EC-GSM device varies from ~350bps to 70kbps, depending on the coverage class currently in use. 

All power classes available for GSM devices are available for EC-GSM. The typical power class used is 33 dBm. An additional lower power class of 23 dBm has also been defined, enabling the integration of the power amplifier onto the chip, enabling longer lifetime and reducing cost in exchange for a short range. Power Saving Mode, which was defined in Release 12, and eDRX (described in NB-IoT section) are also supported on EC-GSM devices, further increasing energy efficiency. In addition, EC-GSM supports a relaxed idle mode behaviour, where no cell measurements are performed while in a Power Saving State \cite{3GPP-tech}.

The battery life of EC-GSM nodes is about 10 years with a 5 Wh battery, depending on several factors including the distance of the device from the base station, the number of bytes required to send per day, and the power class used. Interested readers should refer to tables  [6.2.6.6-9] - [6.2.6.6-12] in \cite{3GPP-tech}. As an example, a device using the 33dBm power class, providing a coverage of 154dB, and sending 50 bytes every 2 hours can be expected to last over 14 years.

\section{Other LPWA solutions}

A number of other LPWA solutions have also been developed, which will be summarised here. Less information is publicly available about these technologies, either because they are proprietary in nature or research focusing on these is still preliminary in nature. 

\textbf{Nwave}'s \cite{Nwave} eponymous protocol is, like Sigfox, based around Ultra Narrow Band communications in the sub 1GHz unlicensed ISM bands. Nwave's nodes can cover 10km in urban environments, and 30km in rural, and can operate for 20 years on a single AA lithium battery, providing a data-rate of 100bps \cite{Nwave-2}. They claim that their advanced de-modulation techniques enable the use of their protocol in these busy bands without the risk of collisions.

\textbf{Ingenu}'s RPMA (Random Phase Multiple Access) \cite{RPMA} protocol is a spread-spectrum solution which operates on the 2.4GHz ISM band. Their use of this band instead of the sub 1GHz bands is based primarily on the relaxed regulations of the band. The coverage RPMA achieves owes mostly to the increased transmission power available at this band - by maximising the transmission power they can achieve a range of 16km. In addition, there are no duty cycle regulations to be followed in Europe in the 2.4GHz band. One deployment of RPMA utilises 1MHz of the 80MHz band - enabling multiple simultaneous deployments or alternatively the use of multiple channels to support one network. 

RPMA uses an adaptive data rate technique, where devices select their optimum spread factor based on the downlink signal strength. The base station is capable of receiving at all spread factors and delay times. In addition devices relay channel conditions inside uplink messages, enabling the base station to optimise the downlink data rate, optimising capacity and energy usage. All messages are encrypted, and a form of the Viterbi algorithm allows the base station to guarantee message arrival even with up to 50\% Packet Error Rate (PER).

\textbf{Telensa} \cite{Telensa} also provide an Ultra Narrow Band solution in the sub 1GHz unlicensed ISM bands. Unlike most LPWA technologies, Telensa claim their protocol can provide fully bi-directional communication - so it is suitable for control as well as monitoring. A Telensa base station can connect to up to 5000 nodes, and cover 2km in urban areas and 4 kms in rural. Individual nodes continue to function as programmed (in smart lighting applications) even if the connection to their base station is lost, and have an estimated lifetime of 20 years \cite{Telensa-2}.

Telensa's solution is the most mature available, the company having been founded in 2005. Telensa have already deployed millions of nodes over 50 smart city networks worldwide, mostly in the United Kingdom but also in cities such as Shanghai, Moscow, and Sao Paulo. The company themselves provide smart lighting and smart parking applications, and also provide a platform that companies can leverage in creating their own smart city applications, enabling authorities to invest in and control the smart city platform for their own city. Telensa is also a member of the Weightless SIG board.

\textbf{Weightless} \cite{Weightless} are a set of LPWA technologies defined and managed by the Weightless-SIG (Special Interest Group). Three different standards have been proposed by the group: Weightless-N, which is focused on ultra-low cost, Weightless-W, which occupies part of the spectrum formerly used by TV whitespace, and Weightless-P, which focuses on high performance. This section will focus on Weightless-P, as it is the most newly defined standard and is most similar to the other LPWA technologies covered in this paper. Like Sigfox, Weightless-P is a narrowband approach on the sub 1GHz ISM bands. Weightless-P splits the spectrum into 12.5kHz channels. Flexible channel assignment, adaptive data rates (from 200bps to 100kbps), and time-synchronised base stations enable the efficient use of spectrum, minimisation of transmit power usage, and prior scheduling of resources, optimising the battery life of individual devices as well as network resources. The support of both full and flexible acknowledgement of all transmissions, Forward Error Correction (FEC), and Automatic Retransmission Request (ARQ) help to maintain reliability and QoS. Weightless-P can support a typical range of 2km in urban environments and all traffic is encrypted using AES-128/256. 

Weightless-SIG is a non-profit standards organisation and this is reflected in Weightless-P design decisions. On the physical layer, standard GMSK and offset-QPSK modulation are used, enabling no dependency on a single hardware vendor. A maximum transmit power of 17dBm further reduces power consumption and enables the use of cheaper integrated power amplifiers, and use of coin cell batteries. Weightless-P devices have $100\mu W$  power usage when inactive. Weightless claim low latency in both uplink and downlink, enabling the support of over-the-air firmware upgrades. Weightless is supported by the core members of the Weightless-SIG: Accenture, ARM, M2COMM, Sony-Europe, and Telensa.

\textbf{WAVIoT} \cite{Waviot} is an Infrastructure-as-a-Service LPWAN-solutions provider from Houston, Texas. Their solution, NB-Fi (Narrowband Fidelity) is a narrow band protocol which communicates on the sub 1GHz ISM subbands. NB-Fi separates the 500 kHz band into 5 thousand channels, and each signal is transmitted in 50 Hz of bandwidth with a minimum bit rate of 50 bod. DBPSK is used as the modulation scheme in the physical layer. WAVIoT gateways can provide -154 dBm of receiver sensitivity, and cover over 1 million nodes. On WAVIoT-developed devices, short bursts of data use 50mA of current, and in idle mode, a few $\mu A$ are used. Devices have a lifetime of up to 20 years, and a 176 dBm link budget. NB-Fi is an open standard, in that WAVIoT will work with interested parties to develop custom devices that utilise the NB-Fi protocol. WAVIoT support three different network types: public, private (city-wide deployment), and enterprise (campus-wide deployment).

NB-Fi operates on a star topology, and can achieve a coverage of over 16 kilometers in an urban environment and over 50 kilometers in a rural environment. The average uplink latency is 30 seconds, and average downlink latency is 60 seconds. NB-Fi is a full stack technology, covering the physical layer up to the application layer. Similarly to Sigfox, data sent through the gateways is stored on a cloud server, and can be accessed from an IoT platform and easily rerouted and manipulated through use of an API.  All data is encrypted bidirectionally from the device to the server using an XTEA 256 bit key.

The \textbf{Dash 7 Alliance Protocol} \cite{Dash7, Dash7-2} (known as Dash7, D7A, or D7AP) is a protocol designed for wireless sensor network applications being developed by the Dash 7 Alliance. D7AP is a full stack protocol, including the application and presentation layers, which functions over the unlicenced sub 1GHz ISM bands. The presentation layer forms a file system; data transmission is in the form of writing to or reading a remote file, and nodes are described with and can be assigned properties, which can be used along with identifiers in the grouping of requests of remote data for different applications. An API is provided to enable interaction with D7AP networks over any interface. 

D7AP networks are formed from endpoints, subcontrollers, and gateways. Gateways remain active continuously, collecting data from endpoints and relaying it back to the server. Subcontrollers have the functionality of gateways but are designed to operate at a lower power and have sleep cycles, their main function being to relay data from endpoints to gateways. In this way D7AP networks utilise a tree topology, or, without the use of subcontrollers, a star. Endpoints can send data directly to a gateway or subcontroller, or alternatively send an all-cast or any-cast, where it waits for acknowledgements from all or at least one gateway respectively. In this way, mobile applications are supported as nodes can communicate with any available gateway. Endpoints also have the ability to send data to each other, and gateways can also query data from endpoints. Endpoints can transmit (asynchronously) to the gateway at any time, and wake up periodically to listen for downlink transmissions. D7AP provides three different defined data rates: 9.6kbps, 55.555kbps, and 166.667kbps, and the distance between endpoints can reach up to 5 kilometers. The modulation scheme used is 2-(G)FSK, PN9 encoding is used for data whitening, and 1/2 FEC encoding is available. The maximum packet size is 256 bytes. Similarly to 802.15.4, AES-CBC is used for authentication and AES-CCM for Authentication and encryption.

\section{Direct Comparison}

\begin{table*}[t]
\centering
\caption{Direct Comparison of LPWA Technologies Part I}

\label{my-label}
\centering
\begin{tabular}{|l|l|l|l|l|l|}
\hline
\textit{\textbf{Technology}}        & \textit{\textbf{LoRa}}            & \textit{\textbf{Sigfox}}    & \textit{\textbf{NB-IoT}} &  \textit{\textbf{EC-GSM-IoT}}    & \textit{\textbf{D7AP}}        \\ \hline
\textit{Topologies supported}       & typically Star, Mesh possible     & Star                        & Star                     & Star                             & Star or tree                  \\ \hline
\textit{Max data rate per terminal} & 50kbps                            & 100bps                      & 60 kbps DL, 50kbps UL    & 70kbps                           & 166.667kbps                   \\ \hline
\textit{Maturity Level}             & Early stages - some deployments   & in use commercially         & early stages             & early stages                     &                               \\ \hline
\textit{Frequency Band}             & sub GHz ISM bands                 & sub GHz ISM bands           & LTE and GSM bands        & GSM bands                        & sub GHz ISM bands             \\ \hline
\textit{MAC Layer}                  & ALOHA-based                       & ALOHA-based                 & LTE-based                & GSM-based                        & CSMA-CA-based                 \\ \hline
\textit{Range/Coverage}             & 2-5km urban, 10-15km rural        & 3-10km urban, 20-50km rural & 164dB                    & 154 - 164dB                      & 0-5km                         \\ \hline
\textit{Founded}                    & 2015                              & 2009                        & 2016                     & 2016                             & 2013                          \\ \hline
\textit{Modulation Technique}       & Spread-Spectrum                   & Ultra-Narrow Band (UNB)     & LTE-based?               & GSM-based                        & 2-(G)FSK                      \\ \hline
\textit{Proprietary aspects}        & Physical layer                    & Physical and MAC layers     & Full stack               & Full stack                       & Open standard                 \\ \hline
\textit{Nodes per gateway}          & \textgreater1,000,000             & \textgreater1,000,000       & 52,000                   & 50,000                           &                               \\ \hline
\textit{Deployment model}           & Private and Operator-based        & Operator-based              & Operator-based           & Operator-based                   & Private                       \\ \hline
\textit{Encryption}                 & AES                               & Not built in                & 3GPP (128-256bit)        & 3GPP (128-256bit)                & AES-CCM                       \\ \hline
\end{tabular}
\end{table*}

\begin{table*}[t]
\centering
\caption{Direct Comparison of LPWA Technologies Part II}
\label{my-label-2}
\begin{tabular}{|l|l|l|l|l|l|}
\hline
\textit{\textbf{Technology}}         & \textit{\textbf{NB-Fi (WAVIoT)}}      & \textit{\textbf{Nwave}} & \textit{\textbf{Telensa}}                             & \textit{\textbf{RPMA (Ingenu)}} & \textit{\textbf{Weightless-P}} \\ \hline
\textit{Topologies supported}        & Star                                  & Star                     & Star                                                  & Star                         & Star                              \\ \hline
\textit{Max data rate per terminal}  & 100bps                                & 100bps                   & ---                                                   & 8kbps                        & 100kbps                           \\ \hline
\textit{Maturity Level}              & in use commercially                   & early stages             & in use commercially (\textgreater9m devices deployed) & in use commercially          & early stages                      \\ \hline
\textit{Frequency Band}              & sub GHz ISM bands                     & sub GHz ISM bands        & sub GHz ISM bands                                     & 2.4 GHz band                 & sub GHz ISM bands                 \\ \hline
\textit{MAC Layer}                   & ---                                   & ---                      & ---                                                   & RPMA-DSSS                    & ---                               \\ \hline
\textit{Range}                       & 16.6km urban, 50km rural                & 10km urban, 30km rural   & 2-3km urban, 4-10km rural                             & 4km                        & 2km urban                         \\ \hline
\textit{Founded}                     & 2011                                  & 2010                     & 2005                                                  & 2008, renamed 2015           & 2012                              \\ \hline
\textit{Modulation Technique}        & Narrowband                            & UNB                      & UNB                                                   & Spread-Spectrum              & Narrowband                        \\ \hline
\textit{Proprietary aspects}         & Full stack                            & Full stack               & Full stack                                            & Full stack                   & Open standard                     \\ \hline
\textit{Nodes per gateway}           & 1,350,000                             & 1,000,000                & 5,000                                                 & 500,000                      & ---                               \\ \hline
\textit{Deployment model}            & Private and operator-based            & Private                  & Private                                               & Private                      & Private                           \\ \hline
\textit{Encryption}                  & XTEA                                  & ---                      & ---                                                   & AES-128bit                   & ---                               \\ \hline
\end{tabular}
\end{table*}

\subsection{Physical Layer}

As summarised in Tables VI and VII, the majority of solutions utilise the unlicenced ISM bands, in Europe in particular the 868MHz band. The band is also used by 802.11ah, Zwave, Zigbee, and many short range devices. There is no overall control of access to the 868MHz band, beyond the duty cycle regulations -- none of the LPWA technologies mentioned in this paper use any sort of LBT system, and some use a simple ALOHA-based access method. For use cases where the device is only required to report back once per day, unless density is very high, there should be no hindrance to performance -- but this does limit the potential density of nodes. There is also as of yet little focus in research on the co-existence of these different networks. In \cite{Reynders}, the authors study the co-existence of LoRa and Sigfox networks, and show that interference does occur when node density is high; that LPWA protocols must take into account their co-existence with different heterogeneous technologies on the same band in order to maintain QoS when the node density is high.

Technologies utilising spread spectrum techniques frequently describe packets sent at the same frequency but different spread factors as orthogonal, but in reality they are not –- the orthogonality of the different factors is a term used to describe simply how packets at different spread factors can be differentiated based on the codes used. The higher the spread factor used, the higher the range but also the more codes used per symbol, and hence the more inefficient the spectrum efficiency is - spread spectrum techniques are inherently poor in terms of spectrum efficiency. In addition, in \cite{Reynders-2} the authors show that spread spectrum techniques, in a dense environment, cannot differentiate between different spread factors in all cases. In \cite{Goursaud} the authors show that inter-spread-factor interference is possible when one signal is received with a significantly higher power. The limited number of channels in spread spectrum-based technologies means that only a limited number of devices can be transmitting simultaneously - two nearby devices cannot use the same spread factor on the same frequency at the same time. If one signal is received at a higher power than the other (at least 6dB) then that signal is received, otherwise both are lost \cite{Goursaud}. The use cases of LPWA would be those that require only a few transmissions per day - but this is still a limitation on scalability. However, spread spectrum techniques generally require cheaper parts than narrowband techniques.

\subsection{MAC Layer}

Both Sigfox and LoRaWAN make use of ALOHA-based MAC layers. As Tables VI and VII indicate, the other non-cellular solutions do not give details on their MAC layers, but since synchronisation packets are generally not required it can be speculated that ALOHA-based MAC layers are used in these as well. ALOHA-based MAC layers are energy efficient, reducing control plane packets, and enabling asynchronous nodes, but they do not scale well -– as the number of nodes trying to transmit simultaneously increases, so does the backoff. On the other hand, Cellular IoT options have dedicated control channels, and a centralised control of the frequency space used, enabling greater management of scalability. 

\subsection{Business Model}

Tables VI and VII show that the majority of LPWA operators offer their technology using a private-network deployment model. Privately deployed networks that follow no regulation beyond following the duty cycle limitations will be difficult to manage. Sigfox's licence-based model will enable the operator to know of all the networks using Sigfox in an area, allowing the management of the number of gateways required and the deployment of new networks. However, physical interference from other technologies in the same band is still a problem. LoRa hardware can be produced and sold at a cheap price, and LoRa devices have no subscription cost, but the LoRa Alliance only has limited control over the deployment of networks. Cellular IoT technologies on the other hand, though the devices will have a subscription charge, the deployment of gateways is for newer grade hardware as simple as applying a software update; a country can have a functioning NB-IoT or EC-GSM-IoT network within hours.

\section{Conclusions}

The unique combination of range and low energy that LPWA technologies provide give them the potential to be used in a variety of IoT applications. In this paper, we provide a definition of the LPWA, and defines a revised list of characteristic criteria for LPWA -- bringing together a few different sources to define a framework for evaluation. The key aspects of LPWA are analysed, and use cases are defined using a systematic approach. Each of the main LPWA technologies currently in development are discussed, compared and contrasted.

Potential avenues for future research in LPWA include further analysis of scalability of different LPWA technologies. The co-existence of technologies in the 868MHz band also merits further study, as well as potential alternative upper-stack layers that make use of LBT-like mechanisms in order to handle dense deployments of nodes. In some use cases, there is not one technology that can provide the ideal characteristics required for transmission at all times, therefore another potential research area of LPWA technologies is in dual-stack systems, in conjunction with other higher energy and higher throughput wireless protocols, where, for example, the LPWA protocol forms the control plane of the system and the higher throughput protocol forms the data plane, or the LPWA technology is used as redundancy to other forms of wireless communication \cite{Margelis} \cite{Dariz}.

In the initial phase of LPWA technology development, the area was dominated by proprietary technologies targeting very particular use cases. Recently we have seen the emergence of more open technologies that target a more general market. We can foresee the arrival of cellular LPWA technologies as the next shift in the space, with in the long term the cellular solutions providing wireless capability to devices with higher QoS requirements, and licence-free solutions being used in the remaining applications.






\section*{Acknowledgment}

This publication has emanated from research conducted with the financial support of Science Foundation Ireland (SFI) and is co-funded under the European Regional Development Fund under Grant Number 13/RC/2077.

\ifCLASSOPTIONcaptionsoff
  \newpage
\fi

\printbibliography



\end{document}